# Multiferroic nematic $d$-wave altermagnetism driven by orbital-order on the honeycomb lattice

Luigi Camerano,[1] Adolfo O. Fumega,[2] Jose L. Lado,[2] Alessandro Stroppa,[3] and Gianni Profeta[1, 3]

[1]*Department of Physical and Chemical Sciences,*
*University of L'Aquila, Via Vetoio, 67100 L'Aquila, Italy*
[2]*Department of Applied Physics, Aalto University, 02150 Espoo, Finland*
[3]*CNR-SPIN L'Aquila, c/o Department of Physical and Chemical Sciences,*
*University of L'Aquila, Via Vetoio, 67100 L'Aquila, Italy*

Altermagnets provide promising platforms for unconventional magnetism, whose controllability would enable a whole new generation of spintronic devices. While a variety of bulk altermagnets have been discovered, altermagnetism in two-dimensional van der Waals materials has remained elusive. Here we demonstrate that the strained honeycomb monolayer $VCl_3$ is an orbital-order-driven ferroelectric altermagnet, exhibiting a significant and switchable spin-splitting. By using low-energy Hamiltonian and first-principles methods in combination with symmetry analysis, we reveal a unique anti-ferro-orbital-antiferromagnetic phase characterized by a 2D nematic $d$-wave altermagnetic spin splitting, tightly coupled with an orbital-ordered induced ferroelectric polarization. Finally, through symmetry mode analysis, we investigate how structural distortions favor the intricate interplay between orbital, altermagnetic, and ferroelectric degrees of freedom. Our study identifies $VCl_3$ as a prototypical 2D orbital-order-driven multiferroic altermagnet on the honeycomb lattice, establishing a van der Waals monolayer featuring altermagnetic ferroelectricity.

Multiferroic materials, defined by the coexistence and the possible coupling of multiple ferroic orders [1], offer a unique platform for functional properties, enabling the control of multiple degrees of freedom (charge, spin, lattice) through external factors. This potential is further enhanced by the remarkable tunability of two-dimensional (2D) van der Waals materials through strain, stacking, and twisting, enabling the creation of exotic phenomena such as unconventional superconductivity [2–4], heavy fermion Kondo systems [5, 6], quantum spin liquid states [7–10], and 2D multiferroicity [11–17]. Altermagnetism has emerged as a fascinating new class of antiferromagnetic order, characterized by non-relativistic spin-splitting of electronic bands in momentum space, even in materials composed of light elements [18–20], while maintaining zero net magnetization. This unique property combines features of both ferromagnetism and antiferromagnetism in a single material, offering unique opportunities for spintronics and quantum materials research. Clearly, the possibility to manipulate the spin properties of an altermagnet with electricity is particularly attractive because electric fields are much easier to manipulate and integrate into modern electronic devices than magnetic fields. Electrical tuning is potentially also faster (subnanosecond) and could use less energy, two crucial properties for the development of high-speed, low-power spintronic devices. Recently, a novel Ferroelectric Switchable Altermagnetism (FSA) has been theoretically proposed, where the reversal of ferroelectric polarization is coupled to the switching of altermagnetic spin splitting [18, 21–23]. Altermagnetism arises in fully compensated magnets where the joint parity and time-reversal ($\mathcal{PT}$) and $\mathcal{T}\tau$, with $\tau$ being the fractional lattice translation, are broken. In such systems, the distinct spin sublattices are linked by rotational or mirror symmetry operations. Since the altermagnetic spin splitting is a non-relativistic property that manifests without the need for spin-orbit coupling (SOC), the symmetry-theoretical framework should adopt spin space group (SSG) [24–26] although alternative theoretical approaches have been also proposed [27, 28]. An interesting case arises when the symmetry breaking is driven not by a structural distortion generating an effective crystal field on the magnetic ions, but rather by electronic degrees of freedom. This phenomenon occurs in orbitally ordered systems, where an electronic instability leads to the formation of inequivalent sublattices with distinct orbital occupancy patterns [17, 29, 30]. The emergence of the orbitally-ordered state can induce the altermagnetic phase via electronic symmetry breaking, unveiling a hidden altermagnetic phase that remains undetectable through conventional spin-space group analysis [31].

The recent discovery of orbital ordering in the 2D magnetic material $VCl_3$ [17] presents an ideal platform for exploring coupled orbital and spin phases in low dimensions. Indeed, $VCl_3$ is part of the vanadium trihalide family, a group of magnetic insulators characterized by their partially filled $d$-shell [32–39]. In this material, electronic degeneracy and strong on-site correlations on $d$-manifold lead to electronic symmetry breaking stabilizing an antiferro-orbital (AFO) ground state coexisting with the magnetic order [17] (see Fig. 1). In contrast with the Ising ferromagnet $VI_3$ [33, 36, 40], where the ferromagnetic coupling is driven by a strong spin-orbit coupling effect [36], $VCl_3$ it is highly tunable by external perturbations such as substrate interactions [41], stacking configurations [32], and strain [17]. The remarkable tunability of $VCl_3$ motivates the exploration of the possible coexistence of an antiferromagnetic (AFM) phase coupled with anti-ferro-orbital (AFO) ordering, potentially giving rise to a novel 2D altermagnet [42, 43]. Using a combination of first-principle calculations, model

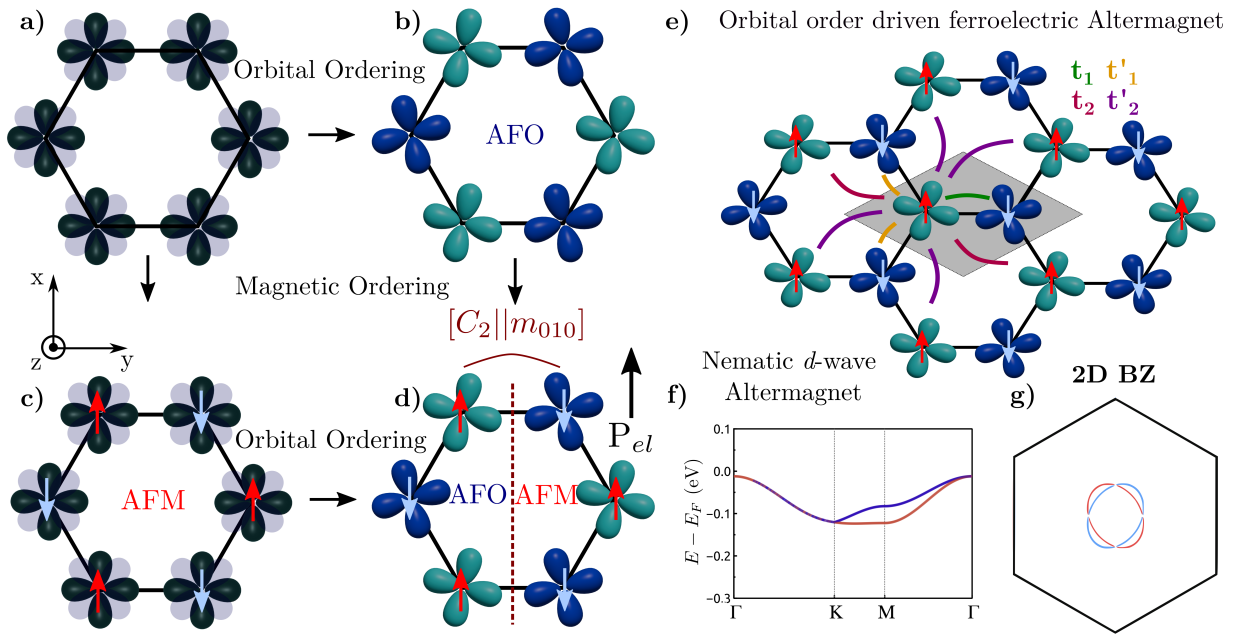

FIG. 1. a) Schematic representation of the high energy phase in honeycomb VCl$_3$, highlighting the not ordered in-plane vanadium $d$-orbitals. b) VCl$_3$ antiferrorbital ordering with blue and tean orbitals highlighting the different orbital configuration. In this phase, crystal symmetry are broken by orbital ordering. c) Magnetic ordering on the not ordered phase. d) Single layer VCl$_3$ showing antiferrorbital-antiferromagnetic ground state. The opposite spin sublattices are connected via mirror symmetry operation $m_{010}$. Combining antiferrorbital ordering with antiferromagnetism give rise to a 2D multiferroic altermagnet. e) Sketch of the two orbital minimal model of AFO-AFM phase with spin-directional hoppings. f) Valence band of the model and in g) energy cut plot in the 2D Brillouin zone highlighting the two nodal lines.

Hamiltonian and symmetry analysis, we demonstrate the emergence of a nematic $d$-wave ferroelectric switchable altermagnetic phase in monolayer VCl$_3$ driven by orbital ordering. The multicomponent order originates from the distinctive structural nature of VX$_3$ compounds characterized by a honeycomb lattice of V cations coordinated by halide ions through edge-sharing distorted octahedra [32–34]. Due to the presence of spin-polarization and orbital degeneracy within a strongly correlated $e'_g$ manifold, the system undergoes spontaneous orbital symmetry breaking, resulting in an orbital-ordered phase (see Fig. 1a-b). The lowest energy state in case of the pristine monolayer is an AFO-ferromagnetic (AFO-FM) phase. We can further tune the magnetic phase by applying strain, inducing an AFO-AFM phase, where compressive strain stabilizes the AFM phase by lowering its energy (see Fig. 1c-d) [17]. In this phase, the two spin sublattices are coupled to the orbital configuration, and, in particular, are connected through a combination of spin flip and mirror operations ($[C_2||m_{010}]$) (see Fig. 1d) [17]. This symmetry creates the conditions for the emergence of an altermagnetic phase. To demonstrate the instrumental role of the AFO-AFM ordering, we construct a minimal tight-binding model on the honeycomb lattice that incorporates all the symmetries to describe the AM phase:

$$H_0 = \beta \sum_i c_i^\dagger (\tau_z \otimes \sigma_z) c_i + \sum_{\langle ij \rangle} t_1(\mathbf{r}_{ij}) c_i^\dagger (\tau_x \otimes \sigma_x) c_i \\ + \sum_{\langle\langle ij \rangle\rangle} t_2(\mathbf{r}_{ij}) c_i^\dagger (\mathrm{I} \otimes \mathrm{I}) c_i + h.c. \quad (1)$$

A schematic representation of this two-orbitals model is shown in Fig. 1e, $\tau$ and $\sigma$ matrices describe the sublattice and spin degree of freedoms respectively. The key ingredient responsible of the altermagnetic behavior lies in the directional dependence of the hopping terms. Specifically, the AFO ordering introduces a directional dependence of the hopping due to the different overlap of the $d$ orbitals as given by the Slater-Koster elements [44]. This leads to inequivalent first ($t_1$ and $t'_1$) and second ($t_2$ and $t'_2$) neighbor hoppings terms (see Fig. 1e). In particular, due to the combination of AFM and AFO orderings, $t_2$ has both spin and directional dependence (the opposite spins have mirrored hoppings), resulting in a momentum-dependent compensated spin-splitting of the energy bands (see Fig. 1f-g) with energy cuts showing two nodal lines.

Our first-principle calculations on VCl$_3$ show that the orbital ordering induces a non-relativistic spin splitting (denoted with $S$) of the bands, as clearly shown in Fig. 2.

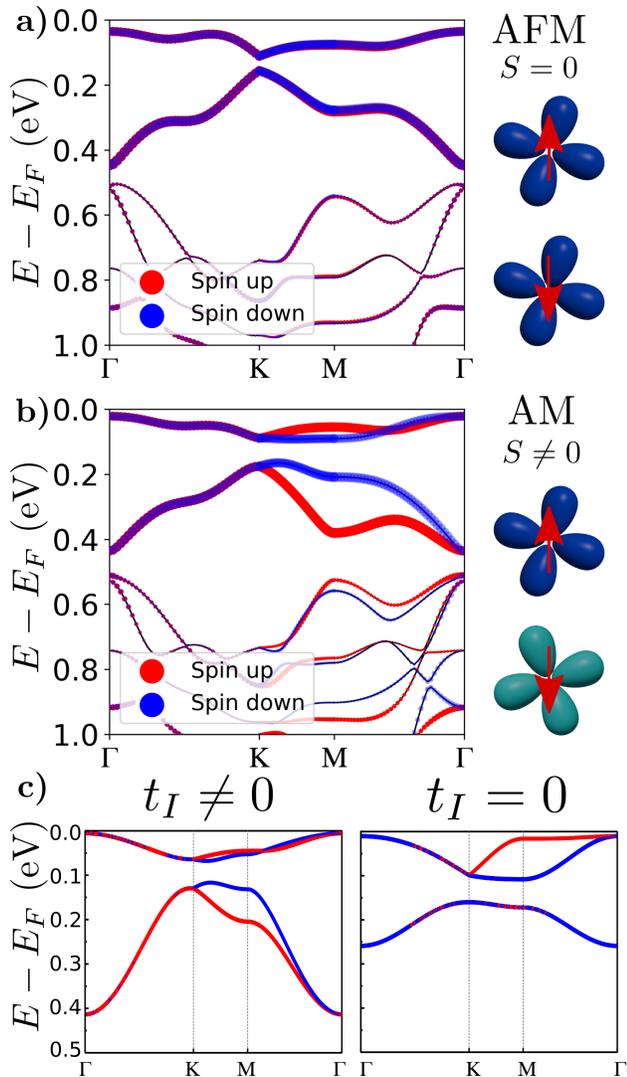

FIG. 2. a) Band structure of the FO-AFM phase, $S$ stands for the AM order parameter. The size of the circles are proportional to the projection of the band structure on vanadiun $d$ orbitals. The Fermi level is fixed a the valence band maximum. b) Band structure of the AFO-AFM phase resulting in a 2D altermagnetic phase with an even parity order parameter. c) Band structure of the four orbital model in presence ($t_I \neq 0$) or in absence ($t_I = 0$) of inter-orbital hopping.

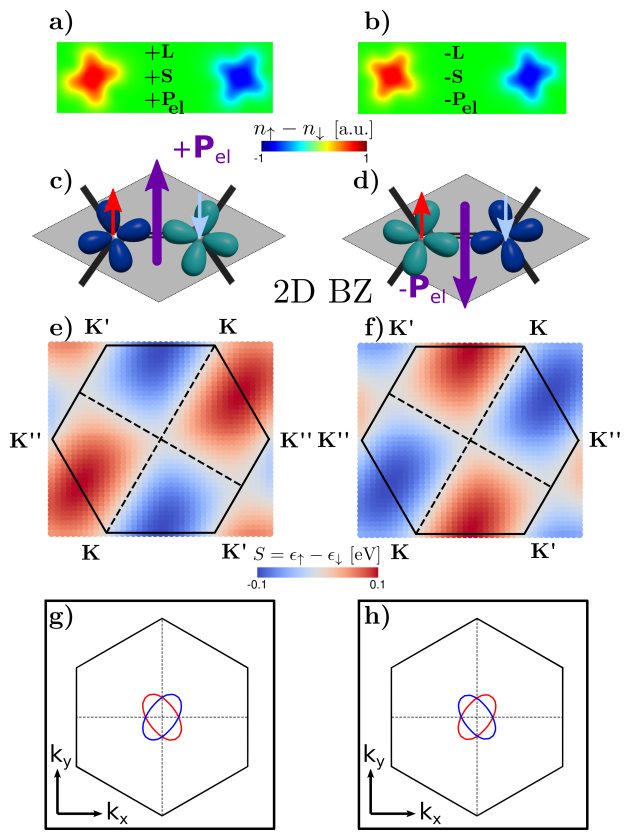

FIG. 3. a) +$L$ electronic configuration. $L$ stands for the orbital configuration, $S$ for the sign of the order parameter and $P_{el}$ is the electronic configuration. b) $-L$ electronic configuration. The arrow point to the direction of the polarization in the $xy$ plane. c) and d) schematics of the different orbital configurations with $P_{el}$ different from zero. e) and f) Corresponding colormap plot of $S$ in the 2D BZ for the V-$d$ conduction state of monolayer VCl$_3$. Different $K$ points are present, signalling the presence of different valley. g) and h) first principle energy cut at the top of the valence band, red for up spin, blue for down spin.

In particular, we show the band structure for two different phases considering the lattice in the symmetric $D_{3d}$ structure: FO-AFM and AFO-AFM. In all the orbitally-ordered (OO) phases the system is a semiconductor with a gap of $E_{gap} \sim 2$ eV consistent with recent experimental findings [32, 41] (see supplementary materials (SM) Fig. S1) with the vanadium $d$-states located at the top of the valence band. The FO ordering (Fig. 2a) combined with an AFM spin state results in an antiferromagnet with degenerate bands due to the presence of time-reversal ($\epsilon_{k,s} \xrightarrow{\mathcal{T}} \epsilon_{-k,-s}$) and inversion symmetry ($\epsilon_{k,s} \xrightarrow{\mathcal{P}} \epsilon_{-k,s}$).

Interestingly, the inversion symmetry breaking [45] induced by AFO ordering in the AFM phase removes degeneracy along some high-symmetry directions of the 2D Brillouin zone (BZ) (see Fig. 2b). The degeneracy along the $\Gamma K$ is protected by the mirror symmetry operation resulting in an even-parity wave anisotropy of the spin-splitted bands. By sampling the all BZ, we find the maximum splitting along the $\Gamma M$ direction with $\Delta E_k^s \sim 0.2$ eV, which is significant compared to other predicted 2D altermagnets [42]. The two vanadium $d$-bands exhibit different characteristics: the lower-energy band has predominantly out-of-plane $a_{1g}$ character, while the higher-energy band at the top of the valence band has in-plane $e_g'$ character. As previously discussed, although only the $e_g'$ manifold drives the OO, spin splitting emerges not only in the $a_{1g}$ band but also in V-$d$ and Cl-$p$ hybridized bands below $-0.5$ eV. This phenomenology can be accounted for by an extended effective model featuring four

orbital, that includes also the $a_{1g}$-$e'_g$ orbitals (see Fig. 2c and SM for a sketch):

$$H = H_0 + \sum_{\langle ij \rangle, l \neq l'} t_I(\mathbf{r}_{ij}) c^\dagger_{il}(\tau_x \otimes \sigma_x) c_{il'} + \mu_{a_{1g}} \sum_i c^\dagger_{i,a_{1g}} c_{i,a_{1g}} \quad (2)$$

where $l, l'$ are the orbitals involved, and $\mu_{a_{1g}}$ is the onsite term for the $a_{1g}$ orbital. In the presence of first neighbor inter-orbital hopping terms, $t_I \neq 0$, spin-splitting also affects the predominantly $a_{1g}$ band, thus accurately capturing the orbital-driven altermagnetic bands observed in the ab-initio calculations. In summary, we have shown how it is possible to stabilize different magnetic orders with different spin splitting driven by orbital and spin order, finding the AFO-AFM phase as an AM.

We now discuss the coupling between electronic and magnetic degrees of freedom, and characterize the symmetries of the AM order parameter. We stabilize two possible orbital configurations, denoted with $+L$ and $-L$, which are characterized by the two distinct AFO orderings (see Fig. 3a-b in which magnetization plots at $z = 0$ are reported). We define $S = \epsilon_\uparrow - \epsilon_\downarrow$ where $\epsilon_{\uparrow(\downarrow)}$ is the single-particle Kohn-Sham eigenvalue for spin up (spin down) and we plot $S$ on a dense $k$ point-mesh ($40 \times 40 \times 1$) in the BZ for V-$d$ conduction band states (see SM Fig. S4 for others V-$d$ states and see Fig. 3e-f). Due to the presence of two nodal lines, we classify VCl$_3$ as a nematic $d$-wave AM on a honeycomb lattice. This should be contrasted with the $i$-wave solution proposed for other 2D AMs with hexagonal BZ [46, 47]. The $C_3$ symmetry of the honeycomb lattice is now lifted due to the electronic symmetry breaking. Additionally, nematicity emerges due to the further reduction of the $C_4$ symmetry, typical of $d$-wave, into $C_2$ symmetry which is protected by the $m_{010}$ mirror plane. Both the symmetry of spin splitting $S$ and of the isoenergy surfaces at the valence band maximum for the $+L$ and $-L$ case respectively (see Fig.3g-h), reveal the topology of the AM nodal surface (see Fig. S5 for energy cuts at different energies showing $C_2$ symmetry). In particular, the sign of $S$ is reversed from $K'$ to $K''$, signaling the emergence of different valleys ($K \neq K' \neq K''$) which can be possibly exploited to induce valley-dependent transport and optical effects [48–50] even without SOC (see SM Fig. S6 for the band structure in the different valley and Fig. S7 for the effect of SOC on the band structure).

So far, we have discussed the interplay between electronic, orbital, and spin degrees of freedom within the centrosymmetric structure. However, our OO states are inherently *polar*. This is confirmed by calculating the electronic polarization $\mathbf{P}_{ele}$ for $+L$ and $-L$ orbital-patterns, which exhibit exactly opposite values of $\pm(1.95, 0, 0.02)$ pC/m (see Fig. S8). The polarization lies in the $xz$ plane as dictated by the $m_{010}$ mirror symmetry. Interestingly, the reversal of $\mathbf{P}_{ele}$ is accompanied by the switching of the OO states and the sign of the spin-splitting $S$ (see Fig. 3), enabling a coupling between spin-polarized bands and electric polarization. This represents a unique realization of a 2D ferroelectric switchable altermagnet [21] achieved without structural symmetry breaking.

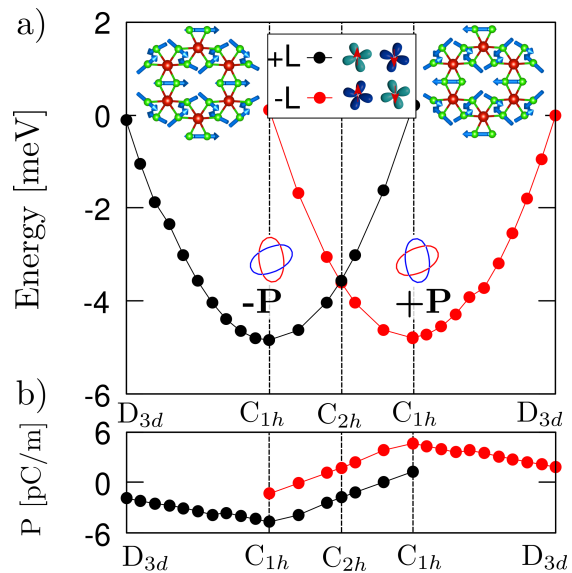

FIG. 4. a) Total energy as a function of the distortion path starting from a $D_{3d}$ symmetry and pointing to the relaxed $+P$ and $-P$ structures. The distortions are reported in the inset. The electronic configuration is fixed at $+L$ (in black) and $-L$ (in red). b) Total polarization as a function of the distortion path.

In presence of orbital degeneracy, many-body effects can produce orbital ordering even in the absence of the ionic contribution; this is known as Kugel-Khomskii mechanism [51] and it is recently revisited and applied to altermagnets [30]. However, for real materials, such as KCuF$_3$ and LaMnO$_3$ a cooperative Jahn-Teller distortion of the nuclei takes place, and thus the Kugel-Khomskii superexchange alone cannot fully account for their physical properties [52, 53]. For this reason, we now analyze the role of possible structural distortions coupled to the orbital ordering ($L$), electronic polarization ($P_{el}$) and non-relativistic spin splitting ($S$) by means of symmetry broken DFT+$U$ calculations [39, 54, 55], which allows us to control the occupation tensor [56] during the structural relaxation. We find that the broken symmetry electronic configuration on the symmetric structure (starting from the FO or AFO phases) leads to a slightly structural distortion $|d| \sim 0.07$ Å of the octhaedra due to the non-zero forces on the ions. This reduces the trigonal symmetry $P\bar{3}1m$ (point group $D_{3d}$) to a monoclinic $C2/m$ (point group $C_{2h}$) in the case of FO ordering, and a monoclinic $Cm$ (point group $C_{1h}$) in the case of AFO ordering (see SM for a table with SSG classification). This demonstrated that the orbital polarization induces the structural distortion rather than being its effect. While the combination of mirror symmetry and two-fold rota-tional axis in $C2/m$ implies centrosymmetry (imposing spin degenerate bands and null polarization), the $Cm$ space group has only a mirror plane symmetry operation generating a nematic $d$-wave altermagnet. Once the $+L$ $(-L)$ is converged in the $D_{3d}$ structure, the system relaxes in the $+P$, $+S$ $(-P, -S)$ following the distortion shown in the inset of Fig. 4. The energy gain associated with structural relaxation is approximately $\Delta E_{rel} \sim 4.5$ meV/per unit cell, very low compared to typical values in other structural distorted ferroelectric materials [21, 57] but comparable with the ones induced by electronic spin-orientation [58]. In contrast with conventional ferroic materials, where structural distortions primarily remove electronic degeneracy, in our case the electronic configuration drives the distortion. Notably, the $+L$ phase can also be stabilized in the $-P,-S$ configuration, albeit at a slightly higher energy (see Fig. 4a) but subsequent ionic relaxation on the $+L,-P, -S$ phase drives the system in the $+L,+P, +S$ phase, without entering the $-L,+P, +S$ phase, further confirming the crucial role of electronic degree of freedom.

Finally, in Fig. 4b we report the polarization as a function of the distortion path. As stated before, in centrosymmetric $D_{3d}$ and $C_{2h}$ structure, there is an electronic contribution to electric polarization, while the ionic one is zero. The relaxation of the ions generates a non zero ionic contribution to the polarization ending up in the relaxed $Cm$ structure where the main contribution is given by P$_{ion}$ (see Fig. S8 in SM for the different contribution to polarization). Interestingly, by a mode decomposition analysis [59–62], we are able disentangle the different contributions to the global distortion and to demonstrate how structural distortion only cause a slight variation in $S$ thereby confirming the electronic origin of the spin splitting (see SM for further details).

In summary, we demonstrated the emergence of multiferroic $d$-wave altermagnetism in monolayer VCl$_3$ due a spontaneous orbital order. Our findings, based on first-principles calculations, a model Hamiltonian, and symmetry analysis, show that the interplay between magnetic, electronic, and structural orders induces orbital ordering, making strained monolayer VCl$_3$ a multiferroic altermagnet. We show that the coupled altermagnetic and ferroelectric orders originate from an electronic symmetry breaking that stabilizes orbital ordering rather than from a structural symmetry as in conventional AM. Our findings unveil a novel framework for identifying 2D altermagnets that are tunable via external factors such as strain and electric fields and simultaneously open new avenues for exploring altermagnetic phases and phenomena in vdW materials.

*Acknowledgements*- G.P. and A.S. acknowledges financial support by the European Union – NextGenerationEU, Project code PE0000021 - CUP B53C22004060006 - "SUPERMOL", "Network 4 Energy Sustainable Transition – NEST" and the European Union - NextGenerationEU under the Italian Ministry of University and Research (MUR) National Innovation Ecosystem grant ECS00000041 - VITALITY - CUP E13C22001060006. J.L.L. and A.O.F. acknowledge the computational resources provided by the Aalto Science-IT project and the financial support from the Academy of Finland Projects Nos. 331342, 358088, and 349696, the Jane and Aatos Erkko Foundation, and the Finnish Quantum Flagship.

# Multiferroic nematic *d*-wave altermagnetism driven by orbital-order on the honeycomb lattice

Luigi Camerano$^a$, Adolfo O. Fumega$^b$, Jose L. Lado$^b$, Alessandro Stroppa $^c$, Gianni Profeta$^{a,c}$

$^a$ Department of Physical and Chemical Sciences, University of L'Aquila, Via Vetoio 67100 L'Aquila, Italy
$^b$ Department of Applied Physics, Aalto University, 02150 Espoo, Finland
$^c$ CNR-SPIN L'Aquila, Via Vetoio, 67100 L'Aquila, Italy c/o Department of Physical and Chemical Sciences, University of L'Aquila, Via Vetoio, 67100 L'Aquila, Italy

## CONTENTS

# I. METHODS

Density functional theory calculations were performed using the Vienna ab-initio Simulation Package (VASP) [1, 2], using both the generalized gradient approximation (GGA), in the Perdew-Burke-Ernzerhof (PBE) parametrization for the exchange-correlation functional [3]. Interactions between electrons and nuclei were described using the projector-augmented wave method. Energy thresholds for the self-consistent calculation was set to $10^{-7}$ eV and force threshold for geometry optimization $10^{-5}$ eV Å$^{-1}$. A plane-wave kinetic energy cutoff of 500 eV was employed for VCl$_3$. The Brillouin zone was sampled using a $12 \times 12 \times 1$ Gamma-centered Monkhorst-Pack grid. To account for the on-site electron-electron correlation we used the GGA+U. The linear response effective Hubbard term $U$ calculated by Ref. [4] is $U = 3.27$ eV. The precondition on the on-site $d$-density matrix was set using the method explained in Refs. [5, 6].

# II. SUPPLEMENTARY FIGURES

## A. Energy gap in monolayer VCl$_3$

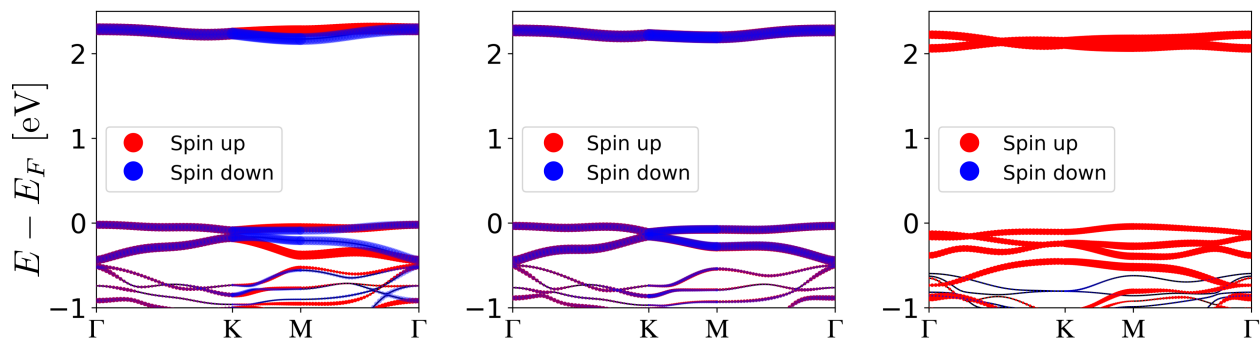

FIG. S1. From left to right: band structure of AM, AFM, FM phases (see main manuscript) showing the gap opening between V-$d$ orbitals. The point size in the plots is proportional to the projection on V-$d$ orbitals

## B. Altermagnetic models on the honeycomb lattice

In this section we will briefly discuss altermagnetic models on the honeycomb lattice. As discussed in the manuscript, the minimal model $H_0$ showing a nematic $d$-wave spin splitting is a two orbital model with spin-directional second nearest neighbors hoppings whose values (in eV) are reported in Tab. S1.

The minimal model can be extended by introducing the Hamiltonian $H$ of the main manuscript, thus including the $a_{1g}$ orbital. The sketch of this four orbital model is reported in Fig. S2. As soon as the inter-orbital hopping is different from zero $t_I \neq 0$, spin-splitting involves also $a_{1g}$ band (see Tab.S2 for the parameters of the model in eV). Note that in these two models orbital ordering determines the directionality dependence of the hoppings. To explicitly include the orbital configuration in the model we build up a minimal four orbital, $H_{OO}$, for the $e'_g$ manifold. Introducing an $\alpha$ parameter that impose the orbital configuration we can write:

$$H_{OO} = \alpha \sum_{i,l,s}(-1)^l \mathbf{c}^\dagger_{ils} \hat{\tau}_z \mathbf{c}_{ils} + \beta \sum_{i,l,s} \sigma^z_{ss} \mathbf{c}^\dagger_{ils} \hat{\tau}_z \mathbf{c}_{ils} + \sum_{ij,lk,s} t(\mathbf{r}_{il,jk}) \mathbf{c}^\dagger_{ils} \left(\hat{\mathbb{I}} + \hat{\tau}_x\right) \mathbf{c}_{jks} + h.c. \qquad (1)$$

where $\tau$ and $\sigma$ matrices describe the sublattice and spin degree of freedoms. A sketch of this model is reported in Fig. S3. Here, by playing with the $\alpha$ parameter both $+L$ and $-L$ configuration can be obtained. For $\alpha = 0$ orbital ordering is quenched and a non-altermagnetic solution arises.

| $t_1$ | $t'_1$ | $t_2$ | $t'_2$ | $\beta$ |
|---|---|---|---|---|
| 0.05 | 0.03 | 0.02 | 0.01 | 0.5 |

TABLE S1. TB parameter for the two orbital model

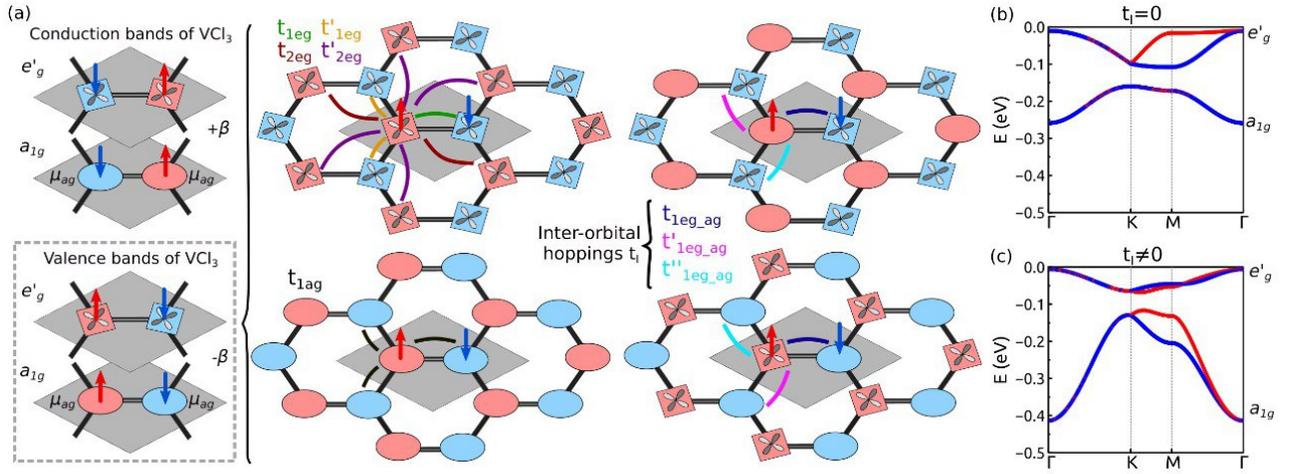

FIG. S2. Sketch of the four orbital model including $a_{1g}$ and $e'_g$ orbitals.

| $t_{1,e'_g}$ | $t'_{1,e'_g}$ | $t_{1,a_{1g}}$ | $t_{1,e'_g a_{1g}}$ | $t'_{1,e'_g a_{1g}}$ | $t''_{1,e'_g a_{1g}}$ | $t_{2,e'_g}$ | $t'_{2,e'_g}$ | $\beta$ | $\mu_{a_{1g}}$ |
|---|---|---|---|---|---|---|---|---|---|
| 0.010 | 0.100 | 0.110 | 0.130 | 0.290 | 0.050 | 0.029 | 0.006 | 0.5 | -0.11 |

TABLE S2. TB parameter for the four orbital $a_{1g}e'_g$ model

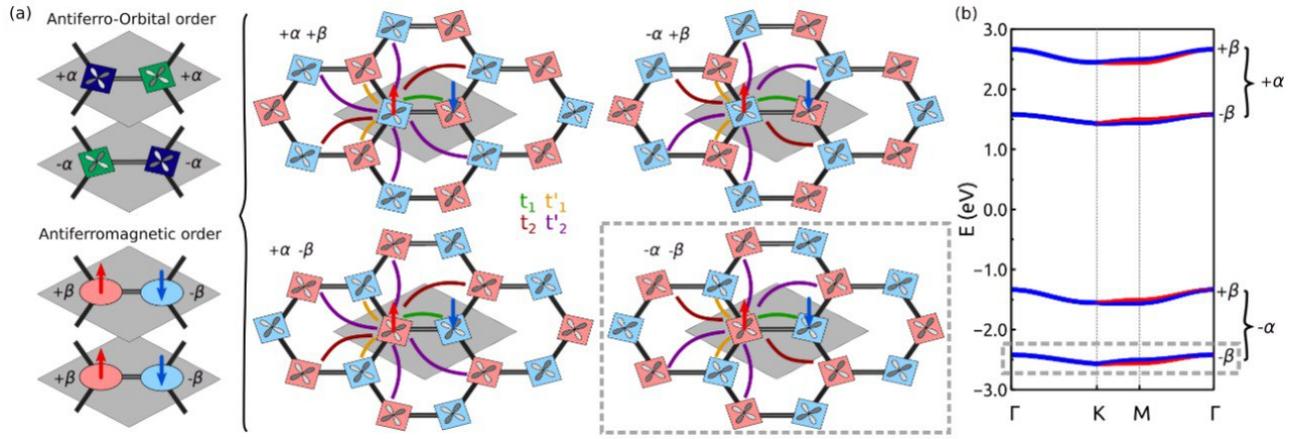

FIG. S3. Sketch of the four orbital model including the $e'_g$ manifold.

C. Symmetry of the altermagnetic spin splitting

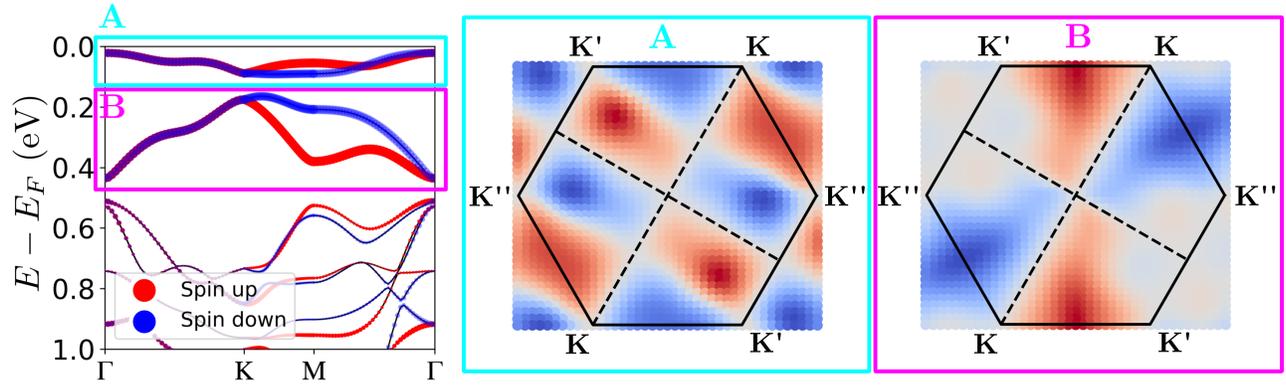

FIG. S4. The defined parameter $S$ (see main manuscript) for different valence bands. In particular, $S$ for the highlighted region (in cyan and magenta) reported in the left panel is reported in center and right panel respectively showing the same $d$-wave symmetry.

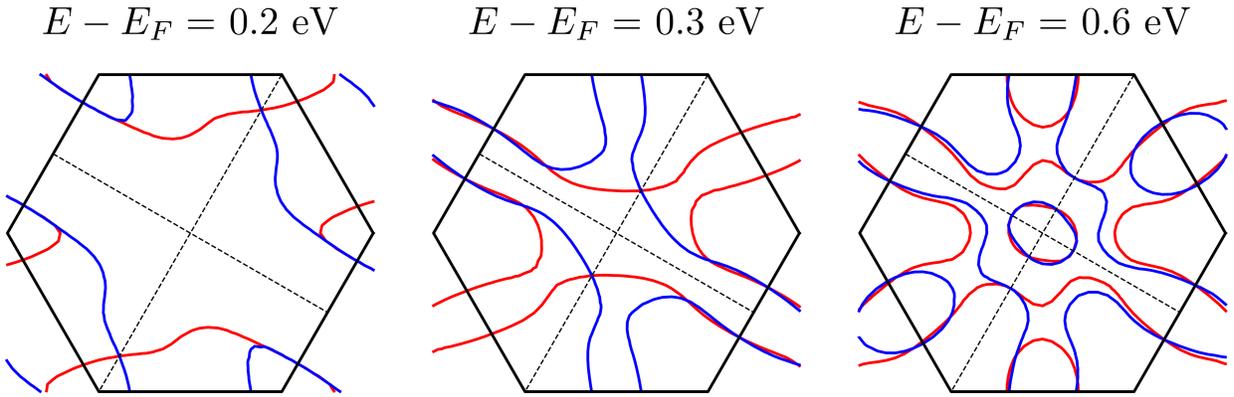

FIG. S5. a) Energy cuts at different energies showing the symmetry of AM order parameter. In particular, it is clear that $C_4$ symmetry is broken resulting in a nematic $d$-wave altermagnet.

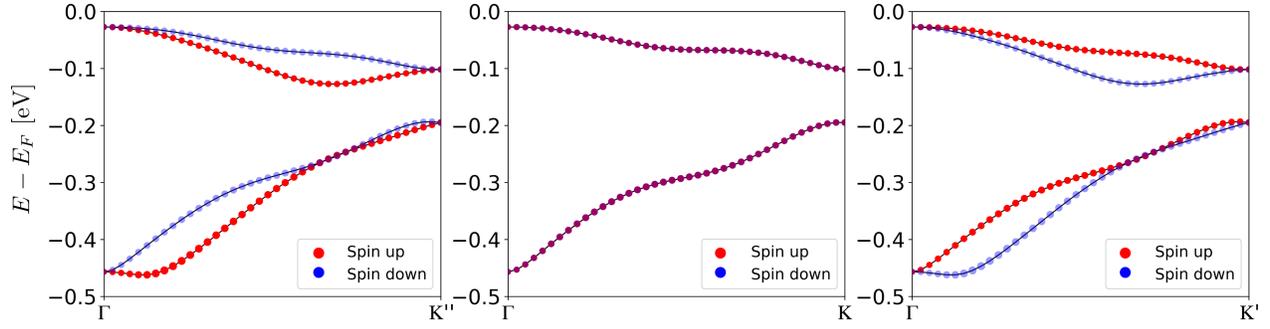

FIG. S6. V-$d$ projected band structure along inequivalent K points in the 2D BZ showing a spin-valley locking in absence of SOC.

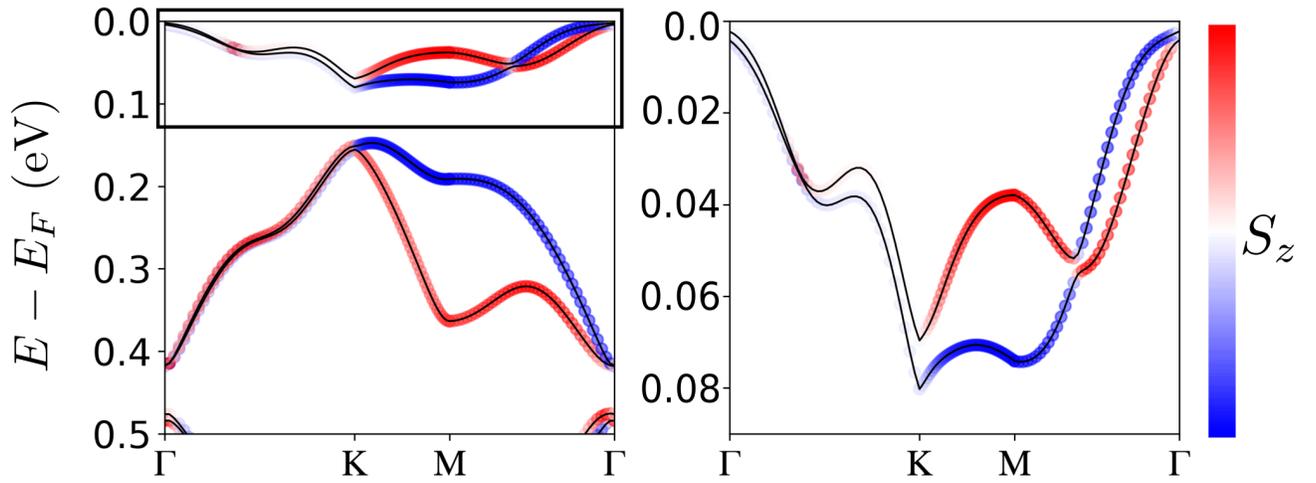

FIG. S7. Effect of spin-orbit coupling on the V-$d$ valence bands showing a minimal modification of the bands.

## D. Polarization and mode decomposition analysis

In this section we use a symmetry mode decomposition analysis (using the Pseudosymmetry [7] and Amplimodes [8] tools of the Bilbao Crystallographic server [9, 10]) to decouple the contribution of different modes on the energy gain, ionic and electronic polarization and spin splitting $S$. Starting from a $P\bar{3}1m$ structure, our symmetry analysis detects 3 modes: the non-polar $\Gamma_3^+$ ($|d| \sim 0.06$ Å) and the polar $\Gamma_3^-$ ($|d| \sim 0.02$ Å) and $\Gamma_2^-$ ($|d| \sim 0.002$ Å). The latter is very small and has not effect on the total energy, polarization and $S$ within our numerical tolerance. In Fig. S7 we report ionic displacement for the $\Gamma_3^+$ and $\Gamma_3^-$, while in Fig. S8 total energy curves show energy gain and the distinct ionic and electronic contribution to the polarization. As expected, while $\Gamma_3^+$ does not contribute to ionic polarization, $\Gamma_3^-$ gives rise to an additional ionic contribution to the polarization.

Regarding the electronic structure configuration, in Fig. S9 we report band structure for the different distortion modes, independently, and the associated $S$ value. Notably, structural distortion has little effect on the energy bands, causing only slight variations in $S$ and thereby confirming the electronic origin of the spin splitting.

| Space Group | Magnetic Space Group | Spin Group | Spin Splitting |
|---|---|---|---|
| $P\bar{3}1m$ | $P\bar{3}'1m$ | $P^{-1}\bar{3}^11^{-1}m^{\infty m}1$ | No |
| $C2/m$ | $C2'/m$ | $C^12/^{-1}m^{\infty m}1$ | No |
| $Cm$ | $Cm$ | $C^{-1}m^{\infty m}1$ | Yes |

TABLE S3. Tables of Space Group, Magnetic Space Group, Spin Group and Spin Splitting for the mode decomposition in monolayer VCl$_3$

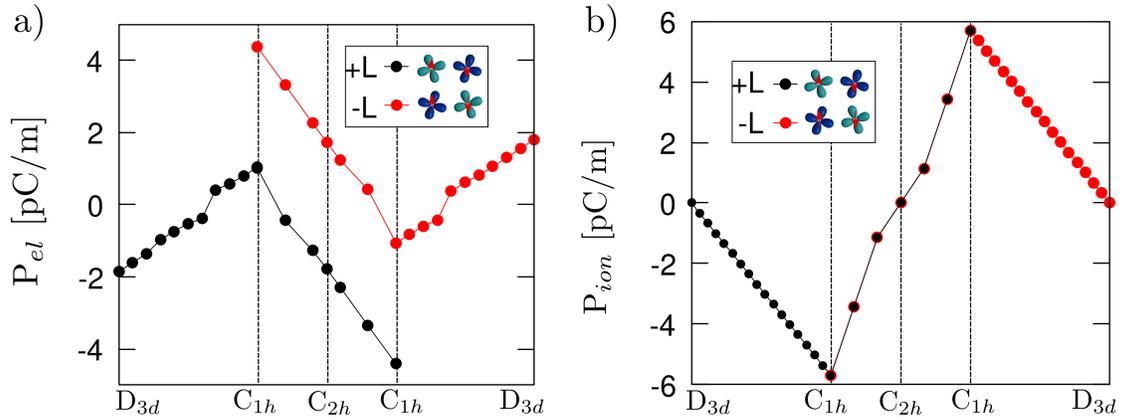

FIG. S8. a) Electronic polarization as $P_{el}$ as a function of the distortion path. b) Ionic polarization as $P_{ion}$ as a function of the distortion path. The polarization for monolayer VCl$_3$ is reported in pC/m following the convention for 2D materials.

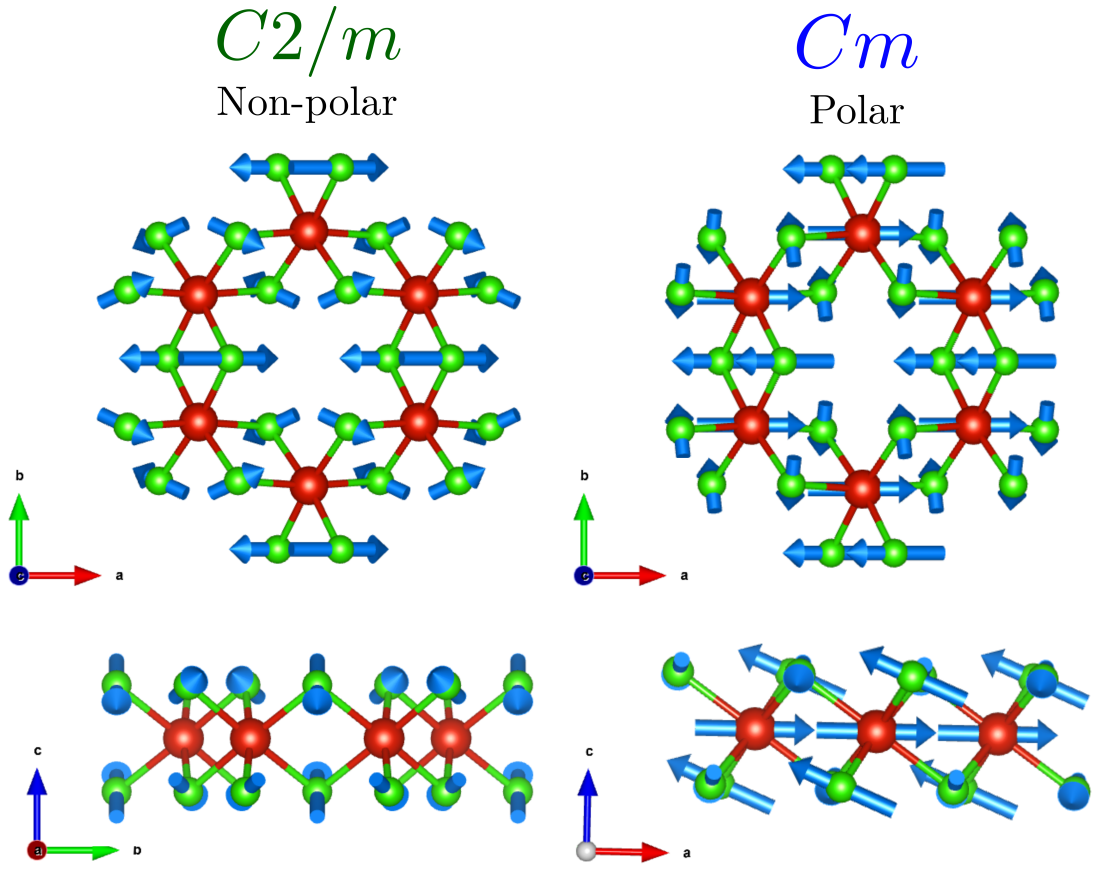

FIG. S9. Left panel: distortion $C2/m$ non-polar mode mainly involving the halides. Right panel: distortive $Cm$ polar mode including both V and Cl.

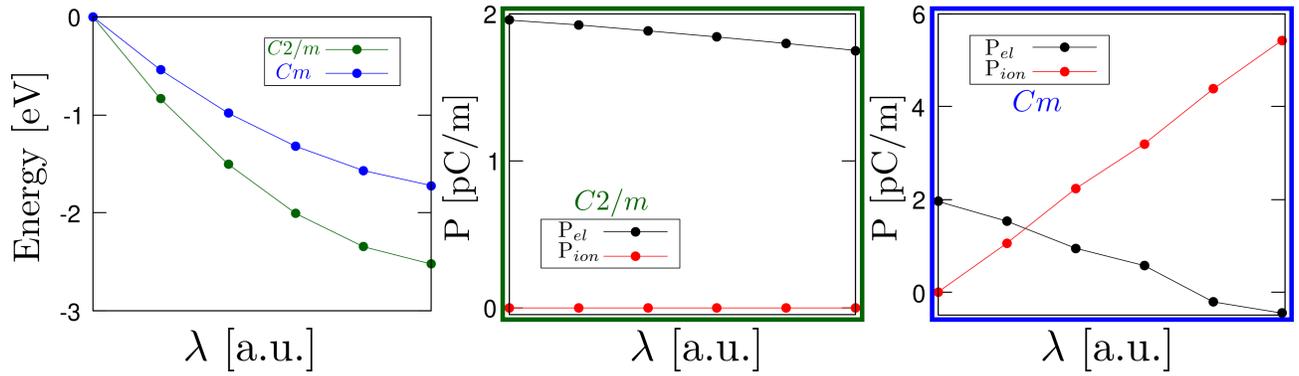

FIG. S10. Left panel: energy as a function of the displacement for the different modes active (see the main text). Following both of the modes decrease the energy of the system. Center panel: electronic and ionic polarization for the $C2/m$ mode. Since it is a non-polar mode, ionic polarization is zero. Right panel: as expected this polar mode activates the ionic polarization. Electronic polarization is different from zero for both the cases.

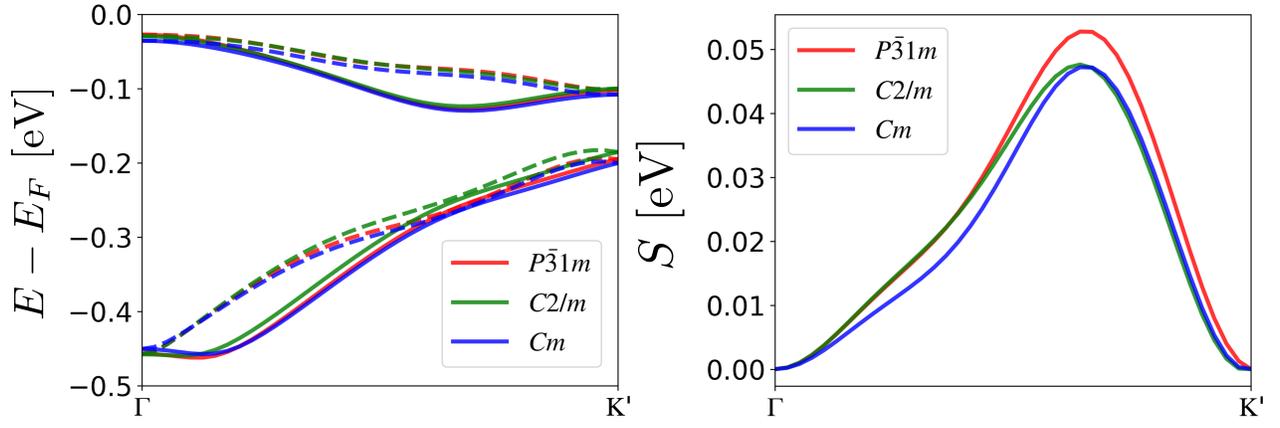

FIG. S11. Left panel: band structure for the differents modes end points. The straght (dashed) line is for spin up (spin down). Right panel: the sign of $S$ along the $\Gamma K'$ line, showing the negligible influence of the distortion modes on the spin-splitting $S$.